\documentclass[12pt]{article}

\usepackage{a4}
\usepackage{xcolor}
\usepackage{graphicx}
\usepackage{epsfig}
\usepackage{amsfonts}
\usepackage{graphicx}
\usepackage{graphicx, epsfig, amssymb} 
\usepackage{bm}
\usepackage{amsfonts}
\usepackage{amsmath}

\usepackage{hyperref}
\begin{document}
\title{Attractive and repulsive Yang-Mills--Higgs  
	\\
	magnetic monopoles on $\R^3$}

\author{ {\large Francisco Navarro-L\'erida}$^{\star}$, 
	{\large Eugen Radu}$^{\diamond}$ 
	and 
	{\large D. H. Tchrakian}$^{\dagger **}$
	\\ 
	\\
	$^{\star}${\small  
		Departamento de F\'isica Te\'orica and IPARCOS, Ciencias F\'isicas,
	}\\
	{\small Universidad Complutense de Madrid, E-28040 Madrid, Spain} 
	\\  
	$^{\diamond}${\small
		Departamento de Matem\'atica da Universidade de Aveiro and }
	\\
	{\small 
		Center for Research and Development in Mathematics and Applications,}
	\\
	{\small 
		Campus de Santiago, 3810-183 Aveiro, Portugal}
	\\ 
	$^{\dagger}${\small 
		School of Theoretical Physics, Dublin Institute for Advanced Studies,}
	\\
	{\small  Burlington Road, Dublin 4, Ireland}
	\\   
	$^{**}${\small Department of Computer Science, National University of Ireland Maynooth, Maynooth, Ireland }
}

\date{}
\newcommand{\dd}{\mbox{d}}
\newcommand{\tr}{\mbox{tr}}
\newcommand{\la}{\lambda}
\newcommand{\ka}{\kappa}
\newcommand{\al}{\alpha}
\newcommand{\ga}{\gamma}
\newcommand{\f}{\phi}
\newcommand{\om}{\omega}
\newcommand{\de}{\delta}
\newcommand{\si}{\sigma}
\newcommand{\vr}{\varrho}
\newcommand{\bomega}{\mbox{\boldmath $\omega$}}
\newcommand{\bsi}{\mbox{\boldmath $\sigma$}}
\newcommand{\bchi}{\mbox{\boldmath $\chi$}}
\newcommand{\bal}{\mbox{\boldmath $\alpha$}}
\newcommand{\bpsi}{\mbox{\boldmath $\psi$}}
\newcommand{\brho}{\mbox{\boldmath $\varrho$}}
\newcommand{\beps}{\mbox{\boldmath $\varepsilon$}}
\newcommand{\bxi}{\mbox{\boldmath $\xi$}}
\newcommand{\bbeta}{\mbox{\boldmath $\beta$}}
\newcommand{\ee}{\end{equation}}
\newcommand{\eea}{\end{eqnarray}}
\newcommand{\be}{\begin{equation}}
\newcommand{\bea}{\begin{eqnarray}}
\newcommand{\ii}{\mbox{i}}
\newcommand{\e}{\mbox{e}}
\newcommand{\pa}{\partial}
\newcommand{\Om}{\Omega}
\newcommand{\vep}{\varepsilon}
\newcommand{\bfph}{{\bf \phi}}
\newcommand{\lm}{\lambda}
\def\theequation{\arabic{equation}}
\renewcommand{\thefootnote}{\fnsymbol{footnote}}
\newcommand{\re}[1]{(\ref{#1})}
\newcommand{\R}{{\rm I \hspace{-0.52ex} R}}
\newcommand{\N}{{\sf N\hspace*{-1.0ex}\rule{0.15ex}%
{1.3ex}\hspace*{1.0ex}}}
\newcommand{\Q}{{\sf Q\hspace*{-1.1ex}\rule{0.15ex}%
{1.5ex}\hspace*{1.1ex}}}
\newcommand{\C}{{\sf C\hspace*{-0.9ex}\rule{0.15ex}%
{1.3ex}\hspace*{0.9ex}}}
\newcommand{\eins}{1\hspace{-0.56ex}{\rm I}}
\renewcommand{\thefootnote}{\arabic{footnote}}

\maketitle
\begin{abstract}
%
An $SO(3)$-gauged Higgs model on $\R^3$ is proposed that, like the Abelian Higgs model on  $\R^2$, features both attractive and repulsive phases, though unlike the latter its solutions do not saturate the topological lower bound.
 What distinguishes this model is that its energy is stabilised by the "Higgs analogue of the Chern-Pontryagin" charge,
 rather than the usual "Higgs--Chern-Pontryagin" charge which is a dimensional descendant of the Chern-Pontryagin charge.
\end{abstract}
\medskip
\medskip


\section{Introduction and motivation}
\label{introduction}
The t'~Hooft-Polyakov (t'HP) monopole~\cite{'tHooft:1974qc,Polyakov:1974ek} is a solution to the equations of motion of the static
Hamiltonian of the Georgi-Glashow (GG) model
\be
{\cal H}_{(\la)}=\frac14\,|F_{ij}^{ab}|^2+\frac12\,|D_i\f^a|^2+\frac18\,\la\,(\eta^2-|\f^a|^2)^2 \,, \label{GG}
\ee
described by the $SO(3)$ gauged Higgs triplet $\f^a$ ($a=1,2,3$) on $\R^3$, where in addition to Yang-Mills (YM) and Higgs quadratic
kinetic terms there appears the symmetry breaking Higgs potential with dimensionless coupling $\la\in[0,\infty)$, and with $\eta$ the VEV
of the Higgs field.

Subsequently, for the system described by  
\re{GG}
 with $\la=0$, $i.e.,$ with no Higgs potential, the explicit
Prasad-Sommerfield~\cite{Prasad:1975kr} (PS) solution was found. 
The most important property of the solutions of the $\la=0$ limit of \re{GG} is,
that they saturate the Bogomol'nyi lower bound~\cite{Bogomolny:1975de}
and solve a system of first-order differential equations, as a result of which these monopoles are non-interacting~\cite{Jaffe:1980mj,Manton:2004tk} by virtue of
the fact that their stress tensor vanishes~\cite{Lohe,Gibbons}.
Inserting a nonvanishing $\la> 0$ {\it by hand},
results in the t'HP monopoles which are exclusively mutually repulsive, see $e.g.,$~\cite{Kleihaus:1998wt}.

A contrasting situation holds for the vortices (solitons) of the Abelian Higgs (AH) model on $\R^2$, whose energy density is
formally given by \re{GG} with the index $a$ running instead over $a=1,2$ $i.e.,$ with the $SO(3)$ 
on $\R^3$ being replaced by $SO(2)$ on $\R^2$, and the triplet Higgs field $\f^a$ replaced by a two component one. 
The stark difference between the GG and AH interpretations of \re{GG} is that in the GG case
the topological stability persists in the limit $\la=0$ while in the AH case the vortex solutions become unstable in this limit
when the topological lower bound vanishes.

The field equations of the AH model can be solved for nonvanishing $\la$,  and for  a particular value of $\la$,
say $\la_{\rm cr}$, for which the first-order Bogomol'nyi equations are satisfied, the vortices are non-interacting. It was shown by
Jacobs and Rebbi (JR) in~\cite{Jacobs:1978ch}, that for $\la>\la_{\rm cr}$ the vortices are mutually repulsive and for $\la<\la_{\rm cr}$
are mutually attractive. 


The search for a formulation of a $3$-dimensional model featuring results, {\rm at least} qualitatively similar to those found by
JR~\cite{Jacobs:1978ch} for the $2$-dimensional AH model, is the purpose of the present work. Previously two such attempts were
made, one in~\cite{Kleihaus:1998gy} and later in~\cite{Grigoriev:2002qc}, in both of which evidence of monopole bound states was 
observed.

The approach used in~\cite{Kleihaus:1998gy} was to employ the $3$-dimensional system
\be
\label{Bu}
{\tilde H}=\tau\eta^2 H(D=3,N=2)+H(D=3,N=3) \,,
\ee
with $\tau$ being a dimensionless coupling constant and $\eta$ with dimension $L^{-1}$.

The term $H(D=3,N=2)$ is just \re{GG} with $\la=0$ which on its own satisfies the Bogomol’nyi-Prasad-Sommerfield (BPS) solutions and is stabilised by the usual
 Higgs--Chern-Pontryagin (HCP) topological charge density~\footnote{In Ref.~\cite{Tchrakian:2010ar}, a hierarchy  of $SO(D)$ gauged $D$-component Higgs models on $\R^D$ was formulated. 
	The energy functions
	of these models, denoted by $H(D,N)$ are stabilised by the topological charges resulting from the dimensional descent of
	Chern-Pontryagin (CP) densities on $\R^D\times S^{2N-D}$ by integrating over the coordinates of the codimension
	sphere $S^{2N-D}$. These (topological) charge densities $\vr(D,N)$ are referred to as HCP densities. The 
	resulting lower bounds are expressed formally by ``Bogomol'nyi inequalities''$H(D,N)\ge\vr(D,N)$.
	
	The ``Bogomol'nyi equations'' that saturate the latter are overdetermined for all models $(D\ge 4,N)$ for all $N$. Only for $D=2$ and
	$D=3$ can these topological lower bounds be saturated. The charge density in the $D=2$ case is that descended from the CP density on
	$R^2\times S^{2N-2}$ which is saturated by the vortices of the AH models~\cite{Arthur:1998nh}. In $D=3$, the charge
	density descends from $\R^3\times S^1$ (only) which is saturated by the PS~\cite{Prasad:1975kr} monopoles.
	In $D=3$, charge densities descending from the CP on $\R^3\times S^{2N-3}$ for $N\ge3$ lead to systems for which the 
	``Bogomol'nyi equations'' are overdetermined and the energies are not saturated. The HCP charge densities $\vr(3,N) $ for $N=3$ and $N=4$ 
	are presented respectively in Sections {\bf 7.3} and {\bf 7.9} of
	Ref.~\cite{Tchrakian:2010ar}, each featuring three independent such parameters $(\la_1,\la_2,\la_3)$, with their numbers increasing with 
	$N$.
	
	An unattractive feature of all (non-Abelian) models $H(D\ge3,N>1)$ is that the desired quadratic kinetic terms $F(2)^2$ and $D\f^2$
	are absent, and only higher order terms appear.
	
}.
The term $H(D=3,N=3)$ in \re{Bu} is the next (more nonlinear) member of the HCP hierarchy of $SO(3)$ gauged Higgs models whose 
energy is bounded by the corresponding HCP charge (see footnote 1). It
is encoded by three dimensionless constants $(\la_1,\la_2,\la_3)$ which play the role that $\la$
plays in the AH model on $\R^2$ albeit in a more complicated and less transparent way, as setting any one of
$\la_n,\;n=1,2,3,$ to zero results in the vanishing of the corresponding topological lower bound~\cite{Jaffe:1980mj}.
Its Bogomoln'nyi-like first-order equations that solve the second-order equations of motion
are overdetermined and result only in the trivial solution.
Moreover this energy density does not feature the usual quadratic kinetic terms $F(2)^2$ and $D\f^2$
which are necessary for ``physical'' reasons and for computational convenience, necessitating the inclusion
of the leading term $H(D=3,N=2)$ in \re{Bu}. Clearly, the model \re{Bu} does not support any solution that saturates
the topological lower bound.
In practice $\tilde H$ defined by \re{Bu} depends on four independent parameters $(\tau,\la_n),\, n=1,2,3,$ rendering the JR type analysis
very complicated in practice. It is relevant at this stage to 
highlight the fact that employing a different topological charge density instead of the HCP charge, leads to a model featuring only one positive 
dimensioness parameter, which is the main purpose of this work.

The other approach to this problem was that adopted in~\cite{Grigoriev:2002qc}, in which the Higgs potential in \re{GG}
is replaced by the $quartic$ Higgs kinetic term coupled with the dimensionless constant $\tau$,
\be
\label{Sut}
\hat H=H(D=3,N=2)+\eta^{-2}\,\tau\,|D_{[i}\f^aD_{j]}\f^b|^2 \,,
\ee
where $(D=3,N=2)$ is the eponymous density appearing in \re{Bu}. Clearly, the energy density
\re{Sut} is bounded below by the corresponding $(D=3,N=2)$ HCP charge density.
Using a numerical annealing method, finite energy lumps with axial and discrete symmetries up to topological charge $8$ were
constructed. This study indicates that monopole bound states are present in that model.

In the studies~\cite{Kleihaus:1998gy} and~\cite{Grigoriev:2002qc}, it was seen that monopole bound states are energetically
viable in some quite different extensions of the PS model.  Both these models employ the HCP topological charge density resulting
from the dimensional descent of some CP density (proposed in~\cite{Tchrakian:2010ar}.)


The aim of the present work is to propose a model on $\R^3$, which like the AH model on $\R^2$ displays a single dimensionless
coupling strength $\ka$. What distinguishes the new model
from those in~\cite{Kleihaus:1998gy} and~\cite{Grigoriev:2002qc} is that the energy density is not bounded by some HCP charge
resulting from the dimensional descent of a CP density, but rather by what we have referred to here as a Higgs--
analogue of Chern-Pontryagin density (HaCP), introduced in~\cite{Matinyan} and presented in detail here in Section {\bf 2}.

The paper is organized as follows.
The new topological charge is introduced in Section {\bf 2}. In Section {\bf 3}, the Bogomol'nyi like lower bounds pertaining to this
new topological charge are exploited to construct the YMH model studied in this work. Symmetry imposition is presented in Section {\bf 4},
 together with the numerical results. A summary of conclusions and some discussion are given in Section {\bf 5}.

\section{HaCP topological charge density for $SO(D)$ Higgs on $\R^D$}

\subsection{General aspects}

To avoid confusion in the nomenclature, and to distinguish the new topological charge from the HCP charges discussed
in~\cite{Tchrakian:2010ar}, we denote the new topological charge as HaCP.
This scheme was first introduced in~\cite{Matinyan}.

The definition of the HaCP density for the $SO(D)$ gauged Higgs model on $\R^D$ starts from the
winding number of the $D$-component Higgs field
$\phi^a$ in $D$ dimensions, which is the volume integral of the density
\be                 
\label{vr0}
\vr_0=\vep_{i_1i_2..i_D}\vep^{a_1a_2..a_D}
\pa_{i_1}\phi^{a_1}\pa_{i_2}\phi^{a_2}..\pa_{i_D}\phi^{a_D}\,.
\ee

In the present context, the real $D$-tuplet Higgs field obeys the usual asymptotics
\be
\label{higgs}
\lim_{r\rightarrow\infty}|\phi^a|^2=\eta^2\, ,
\ee
where $\eta$ is the VEV. Since $\varrho_0$ defined by \re{vr0} is a
total divergence, its volume integral can be evaluated as a {\it surface integral}.

The aim here is to find a density that is both {\it total divergence} and {\it gauge invariant}, and clearly $\vr_0$ defined by
\re{vr0} is not invariant under an $SO(D)$ gauge transformation. To this end, we consider the gauge invariant counterpart of \re{vr0},
where all partial derivatives are replaced by covariant derivatives,
\be                 
\label{vrG}
\vr_G=\vep_{i_1i_2..i_D}\vep^{a_1a_2..a_D}
D_{i_1}\phi^{a_1}D_{i_2}\phi^{a_2}..D_{i_D}\phi^{a_D}\, ,
\ee
which is evidently {\it gauge invariant} but is not a {\it total divergence}.

Our notations for the covariant derivative and curvature are
\bea
D_i\phi^a&=&\pa_i\phi^a+A_i\f^a ,\quad A_i\f^a\stackrel{\rm def.}=A_i^{ab}\f^b \, ,\label{covf}\\
F_{ij}^{ab}&=&\pa_{[i}A_{j]}^{ab}+(A_{[i}A_{j]})^{ab} ,\quad (A_{i}A_{j})^{ab}\stackrel{\rm def.}=A_{i}^{ac}A_{j}^{cb} \, .\label{curv}
\eea
The next step is to calculate the difference $\vr_G-\vr_0$ and cast it in the following form
\be
\label{diff}
\vr_G-\vr_0=\nabla\cdot\Om[A_i,\f^a]-W[F_{ij},D_i\f^a]\, ,
\ee
in which $\Om$ is {\it gauge variant} and $W$ is {\it gauge invariant}.                     

The expression \re{diff} can be calculated straightforwardly using the
Leibniz rule and tensor identities.
This prescription was first introduced in~\cite{Matinyan} for dimensions $D=2,3$.

Two equivalent definitions of the HaCP, $\vr$, follow from \re{diff} as 
\bea
\vr&=&\vr_G+W\label{ginvx}\\
&=&\vr_0+\pa_i\Om_i\,,\label{totdivx}
\eea
of which \re{ginvx} is explicitly gauge invariant and \re{totdivx} is explicitly total divergence, and hence $\vr$ is both
gauge invariant and total divergence. The density
$\vr$ is the topological charge density HaCP which can be employed in establishing Bogomol'nyi like
lower bounds on the energy density.

Explicit expressions of the HaCP topological charge \re{ginvx}-\re{totdivx} for dimensions $D=2,3$ are given in~\cite{Matinyan},
and for dimension $D=4$ in~\cite{Tchrakian:2008zz}. Only in $D=2$ does the definition of the HaCP coincide with that of the
HCP, namely the charge density for the AH vortices.
In all higher dimensions, the definitions of the HaCP completely differ from the HCP in those
dimensions. In particular, the definition of a HaCP in a given dimension is unique, unlike that of the HCP which is encoded with the 
integer $N$ pertaining to the codimension $S^N$ involved in its construction~\cite{Tchrakian:2010ar}.

\subsection{HaCP topological charge density for $SO(3)$ Higgs system on $\R^3$}
In three dimensions, $D=3$, the crucial result \re{diff} is calculated to be
\be
\label{diff3}
\varrho_G-\varrho_0=\frac32\vep_{ijk}\vep^{abc}\left[
\pa_i\left(A_j^{ab}\phi^c\pa_k|\vec\phi|^2\right)-\frac12
F_{ij}^{ab}\phi^c\pa_k|\vec\phi|^2\right]\ ,
\ee
where the notation $|\vec\f|^2=\f^a\f^a$ is used.

It is natural now to define the (gauge invariant) HaCP charge
density as
\bea
\varrho&=&\varrho_G+\frac34\vep_{ijk}\vep^{abc}
F_{ij}^{ab}\phi^c\pa_k|\vec\phi|^2\label{top3i}
\\ &=&
\varrho_0+\frac32\vep_{ijk}\vep^{abc}
\pa_i\left(A_j^{ab}\phi^c\pa_k|\vec\phi|^2\right)\,.\label{top3v}
\eea

This definition is in practice
unique~\footnote{We note here that \re{top3i}-\re{top3v} can be expressed in an equivalent form as
\bea
\varrho&=&\varrho_G+\frac34\vep_{ijk}\vep^{abc}\left(
|\vec\phi|^2\,F_{ij}^{ab}D_k\phi^c\right)\label{top3ix}
\\ &=&
\varrho_0+\frac32\vep_{ijk}\vep^{abc}
\pa_i\left(A_j^{ab}\phi^c\pa_k|\vec\phi|^2-F_{jk}^{ab}|\vec\phi|^2\phi^c\right)\,,\label{top3vx}
\eea
which results by isolating a total divergence term in \re{top3i} which is passed on to the definition of \re{top3v}.

We have eschewed this choice because it leads to the $SO(3)$ curvature term in the energy density of the form
$(|\vec\f|^2\,|F_{ij}^{ab}|^2)$, which is physically undesirable.
},
its main application being the statement of Bogomol'nyi like inequalities for formulating the Higgs model to be
introduced in the next Section. It is the definition \re{top3i} for the HaCP that is exploited for this purpose.

To calculate the topological charge density, it is the definition \re{top3v} of $\vr$ that is employed. The
volume integral of $\vr_0$ is by definition, the {\it winding number}, say $n$. The deviation
of the integral of $\vr$ from $n$ is given by the surface integral
\be
\label{surf}
I=\frac32\int\vep_{ijk}\vep^{abc}
A_j^{ab}\phi^c\pa_k|\phi^c|^2dS_i \, .
\ee

For $I$ to be nonvanishing the integrand in \re{surf} must decay
asymptotically as $r^{-2}$ {\em and no faster}. Now it follows from the finite energy conditions following from the Higgs kinetic terms in \re{H}
that the connection $A_j^{ab}$ decays as $r^{-1}$ by {\em finite energy conditions}, while
\re{higgs}, which enforces the asymptotic constancy of $|\phi^c|^2$,
implies that $\pa_k|\phi^c|^2$ decays as $r^{-(2+\epsilon)}$,
$\epsilon>0$. Thus the integrand in \re{surf} decays as
$r^{-(3+\epsilon)}$ and hence $I=0$.

It follows that the lower bound is given by the volume integral of $\vr_0$ in \re{top3v}, namely by the {\it winding number} of the Higgs scalar.

\section{The topological inequalities and the model}

The Bogomol'nyi like inequalities are
\bea
&&\qquad\qquad\qquad\left|\ka_1^{-1}F_{ij}^{ab}-\frac38\ka_1\,\vep_{ijk}\vep^{abc}\f^c\,\pa_k|\vec\f|^2\right|^2\ge0\,\quad\Rightarrow\nonumber\\
&&\Rightarrow\ka_1^{-2}|F_{ij}^{ab}|^2+\left(\frac34\right)^2\ka_1^2\,|\vec\f|^2\,|(\pa_k|\vec\f|^2)|^2\ge\frac34\,
\vep_{ijk}\vep^{abc}\,F_{ij}^{ab}\f^c\,\pa_k|\vec\f|^2 \, , \label{Bog1}
\eea
and
\bea
&&\qquad\left|\ka_2^{-1}\,D_k\f^c-\frac14\,\ka_2\,\vep_{ijk}\vep^{abc}D_{[i}\f^aD_{j]}\f^b\right|^2\ge0\quad\Rightarrow\nonumber\\\quad
&&\Rightarrow\qquad\ka_2^{-2}\,|D_i\f^a|^2+\frac14\,\ka_2^2\,|D_{[i}\f^aD_{j]}\f^b|^2\ge\vr_G\,,\label{Bog2}
\eea
where, given that the Higgs scalar has dimension $[L^{-1}]$, the constants $\ka_1$ and $\ka_2$ have both dimension $[L]$.

Adding \re{Bog1} and \re{Bog2} yields the Bogomol'nyi like bound \re{top3i}, $\vr$, for the energy density functional
\be
\label{H}
{\cal H}\stackrel{\rm def}=\ka_1^{-2}|F_{ij}^{ab}|^2+\left(\frac34\right)^2\ka_1^2\,|\vec\f|^2\,|(\pa_k|\vec\f|^2)|^2+
\ka_2^{-2}\,|D_i\f^a|^2+\frac14\,\ka_2^2\,|D_{[i}\f^aD_{j]}\f^b|^2\,,
\ee
where the constant $\ka_1^{-2}$ can be identified with the  gluon coupling constant.

What is important about \re{H} is that it features the quadratic kinetic Yang-Mills--Higgs (YMH) terms $F(2)^2$ and $|D_i\f|^2$ as required.

The first-order equations that saturate the lower bound in question are
\bea
F_{ij}^{ab}-\frac38\ka_1^2\,\vep_{ijk}\vep^{abc}\f^c\,\pa_k|\vec\f|^2&=&0 \, , \label{1st1}\\
D_k\f^c-\frac14\,\ka_2^2\,\vep_{ijk}\vep^{abc}D_{[i}\f^aD_{j]}\f^b&=&0\,.\label{1st2}
\eea

With only the two fields $A_i^{ab}$ and $\f^a$, the constraints \re{1st1} and \re{1st2} are (doubly) overdetermined and
are satisfied simultaneously only by the trivial solution.

In what follows, the energy density \re{H} will be rearranged to be expressed in a more transparent form, permitting
a sharper physical insight into it. This is achieved by the following rescaling of the coordinates $x_i$ on $\R^3$
 and the fields 
\be
\label{resc}
x_i\to\la\,\tilde{x_i}\ \Rightarrow\ A_i^{ab}\to\la^{-1}{\tilde A}_i^{ab}\quad {\rm and}\quad \f^a\to\eta\,{\tilde\f}^a \,,
\ee
which would allow expressing \re{H} as
\bea
\label{Hx}
{\cal H}&=&\left(\frac{8}{\la^4\ka_1^{2}}\right)\bigg\{\frac18|\tilde F_{ij}^{ab}|^2
+\frac18\left(\frac34\right)^2\ka_1^4\la^2\eta^6|\tilde\f^a|^2\,|(\tilde\pa_k|\tilde\f^b|^2)|^2\nonumber\\
&&\qquad\qquad\ +
\frac{\ka_1^{2}}{8\ka_2^{2}}\la^2\eta^2\,|\tilde D_i\tilde\f^a|^2+\frac{1}{32}\,\ka_1^2\ka_2^2\eta^4\,|\tilde D_{[i}\tilde\f^a\tilde D_{j]}\tilde\f^b|^2\bigg\}\,.
\eea
It is convenient to choose
$
\la=\frac{2}{\eta}\left(\frac{\ka_2}{\ka_1}\right) 
$
and define
\be
\label{fin}
\ka\stackrel{\rm def.}=\ka_1\ka_2\,\eta^2\, .
\ee
Then,
after 
 suppressing an overall factor $\left(\frac{\eta^4\ka_1^2}{2\,\ka_2^{4}}\right)$ and  relabeling the new coordinates and the fields in \re{resc}
back to $\tilde{x}_i\to x_i$, ${\tilde A}_i^{ab} \to A_i^{ab}$, ${\tilde\f}^a \to \f^a$,
one finds
 the final form of the energy
density function
\bea
\label{tildeH}
{\cal H}=&\left(\frac18|F_{ij}^{ab}|^2+\frac12\,|D_i\f^a|^2\right)\nonumber\\
&&+\ka^2\left[\frac12\left(\frac{3}{4}\right)^2|\f^a|^2\,|(\pa_k|\f^b|^2)|^2
+\frac{1}{32}\,|D_{[i}\f^aD_{j]}\f^b|^2\right]\,,
\eea
which we shall use in what follows.
 
For $\ka=0$, \re{tildeH} reduces to the YMH system \re{GG} without Higgs potential which supports the BPS solutions,
with energy ``equal to'' the topological charge, namely the winding number $n$. Deviations from these BPS solitons arise for nonvanishing values $\ka$.

The total mass-energy of the solutions is computed as the volume integral of the energy
density 
\begin{eqnarray} 
	\label{M}
E=\frac{1}{4\pi}	\int d^3 x \sqrt{{}^{(3)}g}  {\cal H} \, ,
\end{eqnarray}  
with ${}^{(3)}g$ the determinant of the employed 3-dimensional spatial metric
($e.g.$, ${}^{(3)}g=1$ when using Cartesian coordinates).

Finally, let us mention that the solutions satisfy the following 
virial identity
(whose derivation is a 
straightforward  application of the Derrick's scaling argument
\cite{Derrick:1964ww}): 
\begin{eqnarray} 
	\label{Derrick}
	\int d^3 x \sqrt{{}^{(3)}g}\left(
 \frac{1}{8}
  F_{ij}^{ab}|^2 
	+\frac{1}{32}\,\ka^2\,|D_{[i}\f^aD_{j]}\f^b|^2
	\right)
	=
	\int d^3 x \sqrt{{}^{(3)}g} \left(
	\frac{1}{2}\,|D_i\f^a|^2
	+
	\left(\frac34\right)^2\frac{\ka^2}{2}\,|\vec\f|^2\,|(\pa_k|\vec\f|^2)|^2
	\right).
\end{eqnarray}  
Thus the quartic kinetic Higgs term plays the same role as the kinetic YM one,
while the non-standard kinetic Higgs term supplements the usual one, as expected.

\section{The solutions}
\label{results}

 \subsection{The spherically symmetric limit}
 
In the study of spherical and axially symmetric solutions, 
we will employ for numerical convenience  spherical coordinates $(r,\theta,\varphi)$, related to the Cartesian coordinates $(x_1,x_2,x_3)$ by
\bea
\label{sph_coords}
x_1 = r\sin\theta\cos\varphi \, , \ x_2=r\sin\theta\sin\varphi \, , \ x_3=r \cos\theta \, ,
\eea
(and thus $\sqrt{{}^{(3)}g}=r^2 \sin \theta$).

 Subject to spherical symmetry, the $SO(3)$ gauge connection is parametrised by a radial function $w(r)$ as
 \be
 \label{redcon}
 A_i^{ab}=-\left(\frac{1\pm w(r)}{r}\right)\de_i^{[a}\hat{x}^{b]}\ ,\quad \hat{x}^{a}=\frac{x^a}{r}\,,
 \ee
 while the real valued $3$-component Higgs scalar $\f^a$ is written in terms of a radial function $h(r)$,  
 \be
 \label{redhiggs}
 \f^a=
 h(r)\,\hat{x^a}\, .
 \ee
Then
a straightforward computation leads to the following expression of the energy density \re{H} 
of the system
%
\begin{eqnarray}
{\cal H}&=& \frac{1}{r^2} \left[\left(\frac{dw}{dr}\right)^2+\frac{1}{2r^2}(1-w^2)^2\right]
	+ \frac{1}{2}   \left[\left(\frac{dh}{dr}\right)^2+  2\left(\frac{wh}{r}\right)^2\right]
	\nonumber
	\\
	&+&
	\kappa^2\, 
	\Bigg\{
	\frac{9}{8}h^4\,\left(\frac{dh}{dr}\right)^2
	+
	\frac{1}{4 r^2 } 	w^2 h^2 \left[\left(\frac{dh}{dr}\right)^2+\frac12\left(\frac{wh}{r}\right)^2\right]
	\Bigg \} \, .
\end{eqnarray}
The functions $w(r)$ and $h(r)$
solve the equations
\begin{eqnarray} 
	&&
	w''-h^2 w+ \frac{w(1-w^2)}{r^2}
	-\kappa^2 	\frac{wh^2}{4r^2}
	(r^2 h^{'2}+h^2 w^2)=0\, ,\label{eqs-sph1}
	\\
	&&
	\nonumber        
	h''\left[
	1+\kappa^2 \frac{ h^2}{2r^2}
	\left(w^2+ \frac{9}{2}r^2 h^2\right)
	\right]
	+\frac{2h'}{r}
	-\frac{2hw^2}{r^2}
	\\
	&&
	\label{eqs-sph2}        
	\quad
	+ \kappa^2\frac{h}{2r^2}
	\left[
	h'(9 rh^2(h+ rh')+ w(wh'+2 hw'))
	- \frac{h^2 w^4}{r^2}
	\right]=0\, .
\end{eqnarray}
For $\kappa=0$ one can verify the PS solution \cite{Prasad:1975kr}
\begin{eqnarray} 
	\label{1stsol}
	w(r)=\frac{r}{\sinh r},~~h(r)=\coth(r)-\frac{1}{r}\, ,
\end{eqnarray}  
with a mass-energy $E=1$.
For  $\kappa \neq 0$,
the BPS equations (\ref{1st1}), (\ref{1st2})) reduce to the first-order equations
\begin{eqnarray}
	\label{e1}
	w'=0~~{\rm and}~~\kappa  r^2 h^2 h'=\frac{2}{3}(1-w^2)\, ,
\end{eqnarray}
together with  
\begin{eqnarray}
	\label{e2}
	h'=\kappa \frac{h^2 w^2}{2r^2}~~{\rm and}~~w h(1-\frac{1}{2}\kappa h')=0\, .
\end{eqnarray}
However, the system (\ref{e1}), (\ref{e2})
is overdetermined and the  only solution  is the trivial one
\begin{eqnarray}
	w=\pm 1,~~h=0\,.
\end{eqnarray}

The solutions of the second-order ordinary differential equations 
 (\ref{eqs-sph1}), (\ref{eqs-sph2})
are found numerically, by solving a boundary value problem 
(with $w(0)=1$, $h(0)=0$ and  $w(\infty)=0$, $h(\infty)=1$),
$\kappa$ being treated as an input parameter.
The numerical integration was carried out using both the approach described below for axial solutions
 as well 
as the professional solver COLSYS  \cite{Ascher:1979iha}, with  agreement to very high accuracy.

\begin{figure}[h!]
	\begin{center}
		\includegraphics[width=0.5\textwidth]{ 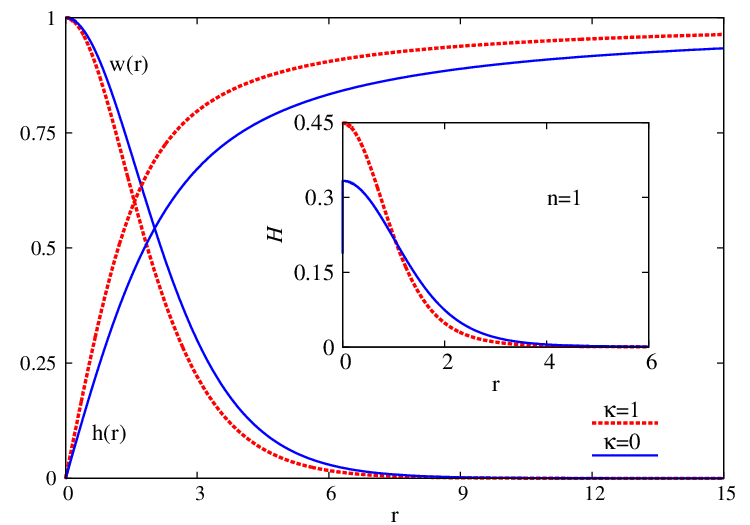}
		\caption{
			The profile of a spherically symmetric solution with
			$\kappa=1$ is shown together with the $\kappa=0$ (BPS) solution.
		}
		\label{profile}
	\end{center}
\end{figure} 

The profile of a typical numerical  solution
is shown in Figure \ref{profile}
together with that of the exact BPS solution (\ref{1stsol}).
As one can see, a nonzero $\kappa$
leads to a more compact profile, while the central value of the
energy density is larger in this case.
A partial understanding of this feature is found when considering
the small-$r$ expansion of the solution, 
\begin{eqnarray} 
	w(r)=1-b r^2+w_4 r^4+\dots\, ,\quad
	h(r)=h_1 r+ h_3 r^3+ \dots \, ,
\end{eqnarray}
where
\begin{eqnarray} 
	w_4=\frac{1}{20}(6b^2+ 2h_1^2 + \kappa^2 h_1^4 )\, ,\quad
	h_3=\frac{bh_1(h_1^2 \kappa^2-4)}{5(2+h_1^2 \kappa^2)}\, ,
\end{eqnarray}
with the central energy density featuring a $\kappa$-contribution,  
${\cal H}(0)=\frac{3}{8}(16 b^2+ 4h_1^2+h_1^4\kappa^2)$.


As expected, the mass-energy of the solutions increases
monotonically with $\kappa$, taking values larger than one, as
seen in the $n=1$ curve in Figure \ref{Mn1}.

\subsection{Axially symmetric monopoles}


In this work we shall restrict ourselves to the axially symmetric solutions.
We will use the Ansatz for the gauge connection and the Higgs field given in Refs. \cite{Kleihaus:1996vi,Hartmann:2000gx}. 
	In spherical coordinates, its non-vanishing components   read\footnote{Our definitions of the $SO(3)$ gauge covariant derivative of $\f^a$ and the curvature of the connection $A_i^{ab}$
	are  \re{covf} and \re{curv}, respectively, with $a=1,2,3$. }
\bea
\nonumber
&&A_\varphi^{12} = -A_\varphi^{21} = -n \sin\theta\left[H_3 \cos\theta - (1 - H_4)\sin\theta\right] \, , 
\\
\nonumber
&&A_r^{13} = -A_r^{31} = - \frac{H_1\cos(n\varphi)}{r} \, , 
\\
\label{A_ansatz}
&&A_\theta^{13} = -A_\theta^{31} = -(1 - H_2)\cos(n\varphi)\, , 
\\
\nonumber
&&A_\varphi^{13} = -A_\varphi^{31} = n\sin\theta \sin(n\varphi) \left[H_3\sin\theta +  (1-H_4) \cos\theta\right] \, , 
\\
\nonumber
&&A_r^{23} = - A_r^{32} = -\frac{H_1 \sin(n\varphi)}{r} \, , 
\\
\nonumber
&&A_\theta^{23} = -A_\theta^{32} = -(1 - H_2)\sin(n\varphi)\, , 
\\
\nonumber
&&A_\varphi^{23} = -A_\varphi^{32} = -n\sin\theta \cos(n\varphi)\left[H_3 \sin\theta +(1 -H_4)\cos\theta\right] \, , 
\eea
for the gauge connection, and
\bea~
\nonumber
&&\phi^1 = \cos(n\varphi)(\Phi_1\sin\theta + \Phi_2\cos\theta) \, , 
\\
\label{A_ansatz2}
&&\phi^2 = \sin(n\varphi)(\Phi_1\sin\theta + \Phi_2\cos\theta) \, ,  
\\
\nonumber
&&\phi^3 = \Phi_1\cos\theta - \Phi_2\sin\theta \, ,
\eea
 for the Higgs field.
	In the above expressions, 
$H_1,H_2,H_3,H_4,\Phi_1$, and $\Phi_2$ are functions of $r$ and $\theta$.
 
 Note that the spherically symmetric case corresponds to setting $n=1$
and $H_1=H_3=0,$ $H_2=H_4=w(r)$, $\Phi_1=h(r)$, $ \Phi_2=0$ in (\ref{A_ansatz}),
(\ref{A_ansatz2}).
A value $n>1$ leads to axially symmetric configurations. 
The  equations for the four gauge potentials and
two Higgs functions are
  solved subject to the following boundary conditions
\begin{eqnarray} 
	H_1=H_3=0,\: H_2=H_4=1,\:\Phi_1=\Phi_2=0  \,	,
\end{eqnarray}
at $r=0$, and
\begin{eqnarray} 
	H_1= H_2=H_3=H_4=0,\:\Phi_1=1,\: \Phi_2=0 \, ,	
\end{eqnarray}
at infinity. 
On the $z$-axis ($\theta=0,\pi$) one imposes
\begin{eqnarray}  
	H_1=H_3=0,\: \partial_\theta H_2=\partial_\theta H_4=0,\:\partial_\theta \Phi_1=\Phi_2=0 \,	.
\end{eqnarray}
In addition, regularity requires $H_2=H_4$ on the $z-$axis.

In the numerics we have used
 a professional elliptic PDE solver \cite{schoen}
based on the Newton-Raphson procedure, employing a
compactified radial coordinate
$x=r/(c+r)$,
with $c$ a properly chosen constant,
usually taken to be one.
The employed software provides various error estimates for each unknown function,
which allows us to judge the quality of the computed solutions.
The numerical error for the solutions reported here is estimated to be typically
$<10^{-4}$.
(Note that the virial identity (\ref{Derrick}) provides another 
error estimate, which is consistent with this value.)

We have considered solutions with $0\leq\kappa \leq 7$
and $n=1,\dots, 8$.
The profiles of  the gauge and Higgs functions,
as well as of the energy density look similar
to those of the    $\kappa =0$ BPS solutions
(although, as for the spherically symmetric case, they become more compact
as $\kappa$ is increased.)

\begin{figure}[h!]
	\begin{center}
		\includegraphics[width=0.45\textwidth]{ 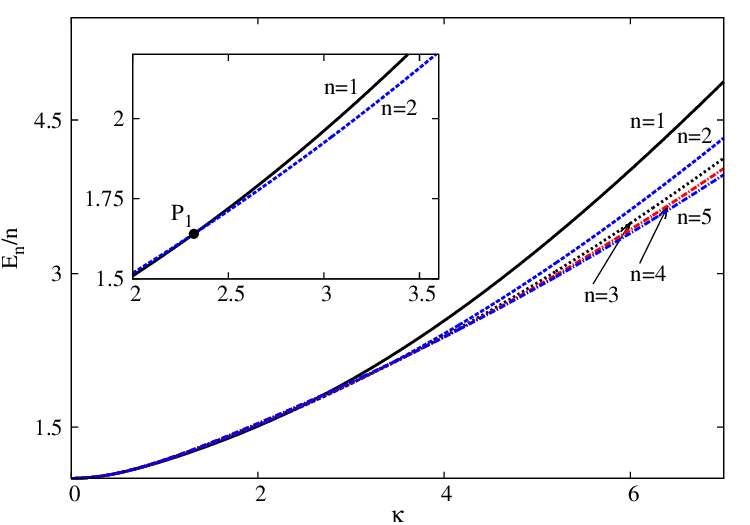}
		\includegraphics[width=0.45\textwidth]{ 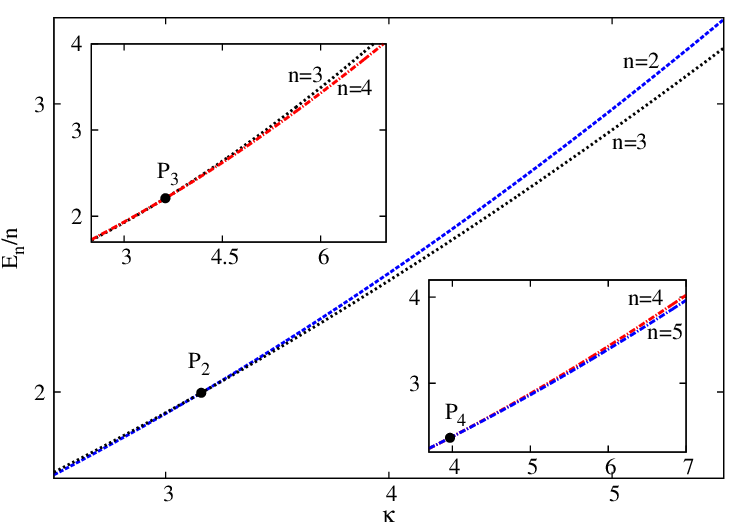}
		\caption{
			The  mass-energy per unit charge is shown as a function of the coupling constant $\kappa$
			for the first five values of $n$.
		}
		\label{Mn1}
	\end{center}
\end{figure} 

\begin{figure}[h!]
	\begin{center}
		\includegraphics[width=0.45\textwidth]{ 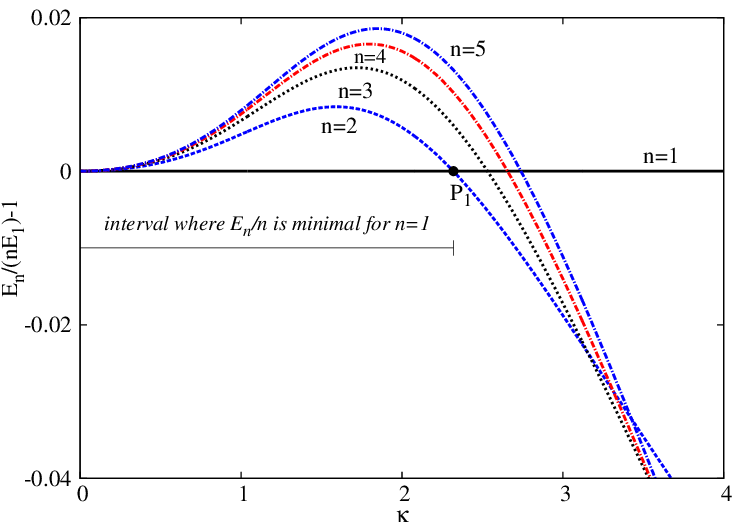}
		\includegraphics[width=0.45\textwidth]{ 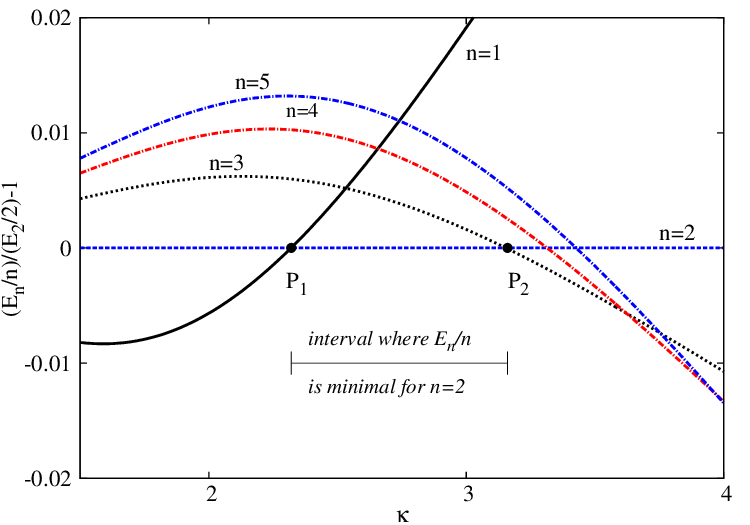}
		\caption{
			The  mass-energy per unit charge 
			divided by the corresponding value for $n=1$ (left) and $n=2$ (right),
			is shown as a function of the coupling constant $\kappa$
			and several values of $n$.
		}
		\label{Mn2}
	\end{center}
\end{figure} 

In Figure \ref{Mn1} we plot the mass-energy per unit charge of the solutions
(as computed according to (\ref{M})), 
shown
	as a function of the coupling constant $\kappa$,
	for $n=1,\dots 5$.
As one can see, the mass-energy always increases with $\kappa$, as expected.	
There one notices the existence of a special set of points $P_k$
where the mass-energy per unit charge curves corresponding to $n=k$ and $n=k+1$ intersect.
That is, among those two sets of solutions,
the mass-energy per unit charge is minimal for  $n=k$ up to $\kappa(P_k)$;
for larger values of $\kappa$, the situation changes and the minimum 
is achieved by the set with $n=k+1$.
One finds 
$\kappa(P_1)=2.3193$,
$\kappa(P_2)=3.1540$,	
$\kappa(P_3)=3.6221$,
$\kappa(P_4)=3.9656$,
$\kappa(P_5)=4.2418$,
$\kappa(P_6)=4.4750$
and
$\kappa(P_7)=4.6780$.

Further insight into this aspect is found in Figure \ref{Mn2}, which is the central result 
of this work.
There we plot the ratio $E_n/n$ divided by the corresponding value for $n=1$ and $n=2$, $i.e.$ $E_1/1$ and $E_2/2$, respectively.   
As one can see, for $0<\kappa<\kappa(P_1)$,
the mass-energy per unit charge is minimized by the spherically symmetric solutions. 
However, the situation changes for larger $\kappa$
and in the interval $\kappa(P_1)<\kappa<\kappa(P_2)$,
the mass-energy per unit charge is minimal for the axially symmetric solutions with $n=2$.

A similar picture is found when considering higher values of $n$ (not shown in that figure).
The following pattern appears to emerge.
For $ \kappa(P_k)<\kappa<\kappa(P_{k+1})$ the solutions with the smallest mass-energy per unit charge are the ones with charge $n=k$. 
This brings us to the result that for $\ka$ in $(0,\ka(P_1))$ the multimonopoles are repulsive, since the most favored configuration from an energy point of view is that of $n=1$. 
However, for $\ka > \ka(P_1)$ the energetically favored configurations are those with $n > 1$,
with $n=2$ for $\ka$ in $(\ka(P_1),\ka(P_2))$, $n=3$ for $\ka$ in $(\ka(P_2),\ka(P_3))$, $etc$. 
 As a consequence, there is attraction for $\ka > \ka(P_1)$: the larger the value of $\ka$ is, the stronger the attraction will be. 
We also mention that no limiting behaviour of the gap between consecutive crossing points $\ka(P_k)$ and
$\ka(P_{k+1})$, as $k$ increases, was observed numerically. 

One can use an analogy with atoms to visualize what the attraction-repulsion pattern of these monopoles is like. $n=1$ monopoles would be the analogues to single atoms, $n=2$ monopoles would be the analogues to diatomic molecules (i.e., union of two atoms),  $n=3$ monopoles would be the analogues to triatomic molecules, and so on. For $\ka$ in $(0,\ka(P_1))$, the energetically favored state of the ``atoms'' is as single atoms, i.e., $n=1$ monopoles are the most stable configuration. For $\ka$ in $(\ka(P_1),\ka(P_2))$, the interaction among ``atoms'' forces them to form diatomic molecules, corresponding to $n=2$ monopoles. As the parameter $\ka$ is increased further, for $\ka$ in $(\ka(P_2),\ka(P_3))$, the attraction becomes stronger and triatomic molecules are favored, that is to say, $n=3$ monopoles are then energetically preferred. And that progressive clumping continues further and further as $\ka$ increases.

\begin{figure}[h!]
	\begin{center}
		\includegraphics[width=0.5\textwidth]{ 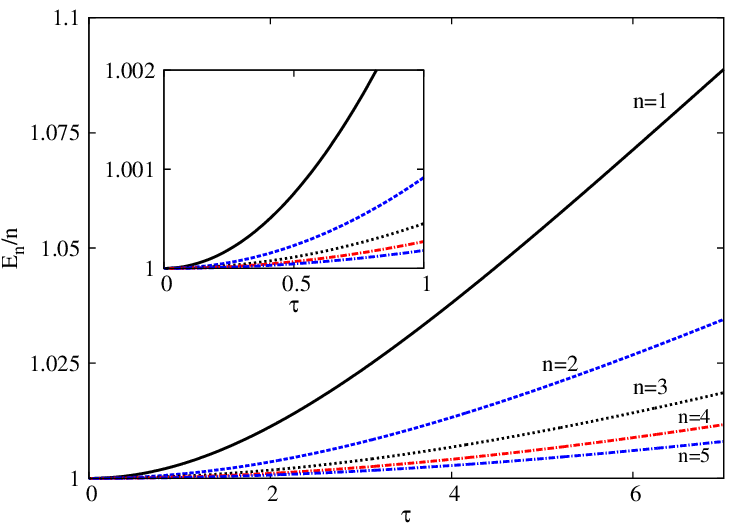}
		\caption{
			The  mass-energy per unit charge is shown as a function of the coupling constant $\tau$
			for several values of $n$.
			Here solutions of the model \re{Sut} are considered (in units with $\eta=1$).
		}
		\label{MnS}
	\end{center}
\end{figure} 
 
 Finally, it is interesting to compare the properties of the magnetic solitons of the model \re{tildeH} whose
 solitons are stabilised by the HaCP charge, and the model \re{Sut} studied in~\cite{Grigoriev:2002qc} whose solitons
 are stabilised by the HCP topological charge. The qualitative difference between the two models is the absence of
 the Higgs kinetic term $|\f^a|^2\,|(\pa_k|\f^b|^2)|^2$ in \re{Sut}.

 To this end, we have integrated the second-order variational equations arising from \re{Sut}, by using the same approach as described above.
  We have found that
 the results are roughly similar to those pertaining to \re{tildeH} (see Figs. \ref{Mn1} and \ref{MnS}).
 Apart from quantitative differences between $(E,\ka)$ and $(E,\tau)$ plots, the striking difference is the absence of any
 crossing points between successive $(E_n/n,E_m/m)$ plots $vs.$ $\tau$ for $n\neq m$, for the model \re{Sut}, that are
 observed in the model \re{tildeH}.
  What both models have in common is that for large enough $\ka$ and $\tau$,
 the magnetic solitons are mutually attractive, implying that this is a result of the dynamics of the
 Higgs quartic kinetic term $|D_{[i}\f^aD_{j]}\f^b|^2$, which in turn implies that mutual attraction is present
 at higher energies.

\section{Summary and discussion}
\label{disc}

The main purpose of this work is to introduce 
an $SO(3)$ gauged triplet-Higgs model  on $\R^3$ which possesses magnetic solitons describing
``monopole bound states''.
The mass-energy of these solutions is bounded below by a topological charge that 
is the integral of what we refer to here as the Higgs analogue Chern-Pontryagin (HaCP) density.

Previously, models supporting monopole bound states were considered in~\cite{Kleihaus:1998gy}
and~\cite{Grigoriev:2002qc},
and in the presence of gravity, in~\cite{Hartmann:2000gx} and~\cite{Gibbons}.
The models studied in~\cite{Kleihaus:1998gy} and~\cite{Grigoriev:2002qc} are based on the application of
the Higgs–Chern-Pontryagin (HCP)~\cite{Tchrakian:2010ar} topological charge, while the model proposed here employs instead the
HaCP~\cite{Matinyan} charge which differs essentially from the HCP charges, except on $\R^2$. Thus, it is that the
classic example of Jacobs and Rebbi~\cite{Jacobs:1978ch} on $\R^2$ is, in this sense, also based on the application
of the HaCP topological charge.

 It may be helpful to recall (alluded to in footnote $[1]$) that the HCP  charges on $\R^D$ presented 
in~\cite{Tchrakian:2010ar}, result from the descent of the Chern-Pontryagin (CP) density on
$\R^D\times S^{2N-D}$ to $\R^D$, and are described by an $SO(D)$ gauged Higgs scalar $\f^a\ ,\ a=1,2,\dots D$.
The most familiar HCP densities are those resulting from the descent of the $N=2$ CP over $S^1$ to $D=3$
supporting Prasad-Sommerfield (PS) monopoles, and those resulting from the $N=2$ CP over $S^2$ to $D=2$
supporting Abelian Higgs (AH) vortices. In both these cases the energy densities resulting from the corresponding
Bogomol'nyi inequalities saturate the topological lower bound. What distinguishes the HCP density characterised
by $(D=3,N=1)$ from that by $(D=2,N=2)$ is that the latter displays $one$ dimensionless parameter $\la$ which
for some critical value $\la_{\rm cr}$ the energy lower bound is saturated~\cite{Jacobs:1978ch}, while
the former does not display any dimensionless coupling and when one is introduced by hand, as in \re{GG}, the
solitons are mutually repulsive~\cite{Kleihaus:1998wt}.
To construct magnetic solitons that have both attractive and repulsive phases, it is necessary that the energy
density be encoded with at least one dimensionless coupling.

One such an attempt was made in~\cite{Kleihaus:1998gy},
employing the HCP density characterised
by the descent $(D=3,N=3)$, which displays $three$ such couplings $(\la_1,\la_2,\la_3)$. Both  attractive and
repulsive phases were indeed observed in the resulting model, but these were not as transparent to study
as in the case of the AH system where only $one$ such coupling was present.

Another attempt at constructing mutually attractive monopoles was made in~\cite{Grigoriev:2002qc}, where the
Higgs potential in \re{GG} was replaced by the Higgs quartic kinetic term in the energy density, \re{Sut}.
As such, this model is also based on the application of a HCP topological charge, namely that applied in
the construction of the Bogomol’nyi-
Prasad-Sommerfield (BPS) and t’ Hooft-Polyakov 
 monopoles.
That system supported monopole bound states of both discrete and
axial symmetry, with the axially symmetric ones having lower masses than those with discrete symmetry, for
a given charge. The mass-energy per unit charge $vs.$ coupling constant profiles were given in Figure \ref{MnS} in the previous section,
where it is shown that these solutions are exclusively attractive.

Finally there is the model \re{tildeH} studied here, which in addition to the usual Yang-Mills (YM) and Higgs quadratic
kinetic terms features a brace of $L^{-8}$ dimension Higgs kinetic terms both with coupling $\ka$. This model emerges
from the application of Bogomol'nyi like inequalities based on the definition of the HaCP topological charge density
introduced in Section {\bf 2}. It features a single positive and dimensionless parameter $\ka$,
in analogy with $\la$ in the AH model.
The plots of mass-energy per unit charge $(E_n/n)$ $vs.$ the dimensionless coupling $\ka$ for different $n$ cross
each other at different values of $\ka$ in contrast with the Abelian case on $\R^2$ where 
all plots $(E_n/n)$ $vs.$ $\la$ cross at the same (critical) value $\la_{cr}$, as all charge $n$ vortices
saturate the Bogomol'nyi lower bound.
By contrast solitons of the model \re{tildeH} on $\R^3$
stabilised by the HaCP charge do not saturate the HaCP lower bound except for $\ka=0$ where the soliton is
precisely the BPS charge-$n$ monopole.
For $\ka>0$ the solutions to the second-order equations with different values of the charge $n$ cross at
different values of $\ka$ in the $(E_n/n)$ $vs.$ $\ka$ plots, as seen in Figures \ref{Mn1} and \ref{Mn2}, with the distance between consecutive crossing points
decreasing with increasing $\ka$. No limiting value of these crossing points was observed numerically.

Like the model~\re{Bu}, the model~\re{tildeH}
describes both attractive and repulsive phases, but unlike \re{Bu}, \re{tildeH} is parametrised by a single
dimensionless coupling, enabling
it to describe a clear cut quantitative display of attractive and repulsive domains (separated by the point $\ka(P_2)$). 
Apart from its simplicity, what makes the model \re{tildeH} special is that in dimension $D=3$ it is unique,
while model~\re{Bu} is only one of infinitely many such models resulting from descents of $(D=3,N)$ for all $N$
rather than $(D=3,N=3)$, that results in the model~\re{Bu}.

As for the overlap of models \re{Sut} and \re{tildeH}, both feature the quartic kinetic Higgs term
$|D_{[i}\f^aD_{j]}\f^b|^2$ but \re{tildeH} in addition features the unusual Higgs kinetic term
$|\f^a|^2\,|(\pa_k|\f^b|^2)|^2$ which is presumably responsible for the mutually repulsive domains in Figures
\ref{Mn1} and \ref{Mn2} in contrast to Figure \ref{MnS}
that describes exclusively attractive solitons.

As a final remark, we would like to point out that all of the solutions discussed in this paper are localized at the origin ($r=0$). As a consequence, the results presented here refer to the asymptotic interaction among monopoles, where the individual monopoles are located at the same point. A complete analysis of the actual attraction/repulsion forces among monopoles separated for finite distances would require to perform fully 3D simulations, which is beyond the scope of this work.

\section*{Acknowledgement}

We thank Burkhard Kleihaus and Paul Sutcliffe for helpful remarks.
F. Navarro-L\'erida gratefully acknowledges support  from MICINN under project PID2021-125617NB-I00 ``QuasiMode".
The work of  E. Radu is  supported by 
CIDMA (\url{https://ror.org/05pm2mw36}) under the Portuguese Foundation for Science and Technology (FCT, \url{https://ror.org/ 00snfqn58}), Grants UID/04106/2025 (\url{https://doi.org/10.54499/UID/ 04106/2025}) and UID/PRR/04106/2025   \newline
(\url{https://doi.org/10.54499/UID/PRR/ 04106/2025}), as well as the projects: Horizon Europe staff exchange (SE) programme HORIZON-MSCA2021-SE-01 Grant No.\ NewFunFiCO-101086251;  2022.04560.PTDC
(\url{https://doi.org/10.54499/2022.04560.PTDC}) and 2024.05617.CERN   \newline
(\url{https://doi.org/10.54499/2024.05617.CERN}).

\begin{small}

\end{small}
\medskip
\medskip

\end{document}